\documentclass[useAMS,usenatbib]{mn2e}

\DeclareSymbolFont{cmletters}{OML}{cmm}{m}{it}
\DeclareMathSymbol{v}{\mathalpha}{cmletters}{"76}

\usepackage{amsmath}
\usepackage{amssymb}
\usepackage{epsfig}
\usepackage{graphicx}
\usepackage{ifthen}
\usepackage{latexsym}
\usepackage{rotating}
\usepackage{subfigure}
\usepackage{times,epsf}
\usepackage{txfonts}
\usepackage{varioref}
\usepackage{verbatim}
\usepackage{url}
\usepackage{color}
\usepackage[dvipsnames]{xcolor}
\usepackage[T1]{fontenc}
\usepackage{aecompl}
\usepackage{float}

\newcommand{\be}{\begin{equation}}
\newcommand{\ee}{\end{equation}}
\newcommand{\bea}{\begin{eqnarray}}
\newcommand{\eea}{\end{eqnarray}}

\newcommand{\figpath}{.}

\newcommand\apj{Astrophysical Journal}

\newcommand\apjs{Astrophysical Journal Suppl. Ser.}
\newcommand\apss{Astrophysics \& Space Science}

\newcommand\aap{Astronomy \& Astrophysics}

\newcommand\nat{Nature}

\newcommand\mnras{Monthly Notices of the Royal Astronomical Society}

\newcommand\pasj{Publications of the Astronomical Society of Japan}

\def\pm{P_\mathrm{mag}}
\def\pg{P_\mathrm{gas}}
\def\pmpg{P_\mathrm{mag}/P_\mathrm{gas}}

\title[Decay of Strong Magnetization]{On the Decay of Strong Magnetization in Global Disc Simulations with Toroidal Fields}
\author[P. C. Fragile \& A. S\k{a}dowski]
       {P. Chris Fragile$^1$\thanks{E-mail: fragilep@cofc.edu}
       and
	Aleksander S\k{a}dowski$^2$\\
        $^1$ Department of Physics and Astronomy, College of Charleston, Charleston, SC 29424, USA\\
        $^2$ MIT Kavli Institute for Astrophysics and Space Research, 77 Massachusetts Ave, Cambridge, MA 02139, USA
}

\begin{document}

\maketitle

\label{firstpage}

\begin{abstract}
Strong magnetization in accretion discs could resolve a number of outstanding issues related to stability and state transitions in low-mass X-ray binaries.  However, it is unclear how real discs become strongly magnetized and, even if they do, whether they can remain in such a state.  In this paper, we address the latter issue through a pair of global disc simulations.  Here, we only consider cases of initially purely toroidal magnetic fields contained entirely within a compact torus.  We find that, over only a few tens of orbital periods, the magnetization of an initially strongly magnetized disc, $\pmpg \ge 10$, drops to $\lesssim 0.1$, similar to the steady-state value reached in initially weakly magnetized discs.  This is consistent with recent shearing box simulations with initially strong toroidal fields, the robust conclusion being that strongly magnetized toroidal fields can not be locally self-sustaining. These results appear to leave net poloidal flux or extended radial fields as the only avenues for establishing strongly magnetized discs, ruling out the thermal collapse scenario.
\end{abstract}


\section{Introduction}
\label{s.introduction}

The importance of magnetic fields to the process of accretion is now widely accepted.  Magnetic fields are paramount to the operation of the magneto-rotational instability \citep[MRI;][]{Balbus91}, which drives turbulence within differentially rotating discs.  This turbulence acts as an effective viscosity, transporting angular momentum and dissipating energy to facilitate accretion onto the central object.  Magnetic fields also likely play a critical role in powering the narrow, fast jets of outflowing gas often associated with discs \citep{Blandford77}.  These fields may also launch wider, slower winds \citep{Blandford82}, observed in multiple black hole low-mass X-ray binaries \citep[LMXBs;][]{Miller06,Ponti12}.

Magnetic fields may also play a more subtle role in helping stabilize accretion discs.  For example, \citet{Begelman07} \citep[also][]{Oda09} argued that strong magnetic fields can stabilize thin discs against both thermal and viscous instabilities.  There are now numerical results to back up that claim \citep{Sadowski16}.  This is potentially very important as it may help explain why accretion discs in X-ray binaries generally show far less variability at high luminosities than would be expected if these instabilities were at play.

So far, models that advocate for strong magnetic fields in accretion discs have failed to demonstrate conclusively how those fields get there in the first place and whether or not they can remain strong.  For instance, discs threaded with a net vertical magnetic flux have been shown in shearing box numerical simulations to achieve high magnetization \citep[$\pmpg \ge 1$;][]{Bai13,Salvesen16a}.  However, it is not clear in all cases that an adequate supply of such flux exists \citep{Sadowski16b}.

Another suggested avenue for achieving high magnetization is through a thermal runaway \citep{Pariev03, Machida06,Fragile09}.  Rapid cooling of an initially hot, thick, weakly magnetized disc can increase the magnetic flux by trapping magnetic field within the collapsing disc.  Yet, even if such a thermal runaway were to occur, it would likely only produce a transient high-magnetization phase, unless there is a mechanism that would allow the disc to persist in such a state.

A third possibility was suggested by \citet{Johansen08}, who proposed a dynamo process mediated by the Parker instability, in which buoyantly rising azimuthal field is replenished by the stretching of radial field created as gas particles slide down inclined field lines.  This dynamo was predicted to maintain high disc magnetizations ($\pmpg \gtrsim 1$), and the authors found support for this claim through shearing box simulations.  However, as pointed out by \citet{Salvesen16}, those simulations used periodic boundary conditions on the top and bottom of the box, thereby preventing magnetic field from escaping.  Shearing box simulations with more realistic ``open'' vertical boundary conditions, on the other hand, found only weak magnetizations \citep[$\pmpg \lesssim 0.1$;][]{Salvesen16}, bringing the Parker-driven dynamo into question.  

In this work, we report on simulations that are specifically designed to test the question of whether discs are able to remain strongly magnetized, assuming they are able to somehow achieve this state.  Similar to \citet{Johansen08} and \citet{Salvesen16}, we focus on the case of initially purely toroidal field configurations.  However, unlike those works, we perform {\em global} simulations.  Three important advantages of global simulations are: 1) there are no concerns about about boundary issues since the domain is larger than the entire disc; 2) any global instabilities or correlations with length scales larger than a typical shearing box will be captured; and 3) global simulations naturally include the curvature terms neglected in shearing boxes.  Of course, two significant disadvantage of global simulations are: 1) the computational cost of achieving adequate resolution; and 2) the limited duration allowed when the simulation is initiated from a finite equilibrium distribution, as in our study.  For example, by the time we stop our strongly magnetized simulation, the disc has lost over 80\% of its initial mass into the black hole.  Nevertheless, we are able to clearly demonstrate that the magnetization of such discs becomes weak on the timescale of a few tens of orbits.

In the next section, we briefly describe our numerical set up.  Then, in Section \ref{s.results} we present our results.  In Section \ref{s.discuss}, we end with some concluding thoughts.  Throughout most of the text, we use units where $G=c=1$, so physical lengths and times are recovered by multiplying by $GM/c^2$ and $GM/c^3$, respectively.

\section{Numerical Set Up}
\label{s.setup}

\begin{figure}
\centering
\includegraphics[width=1\columnwidth]{\figpath/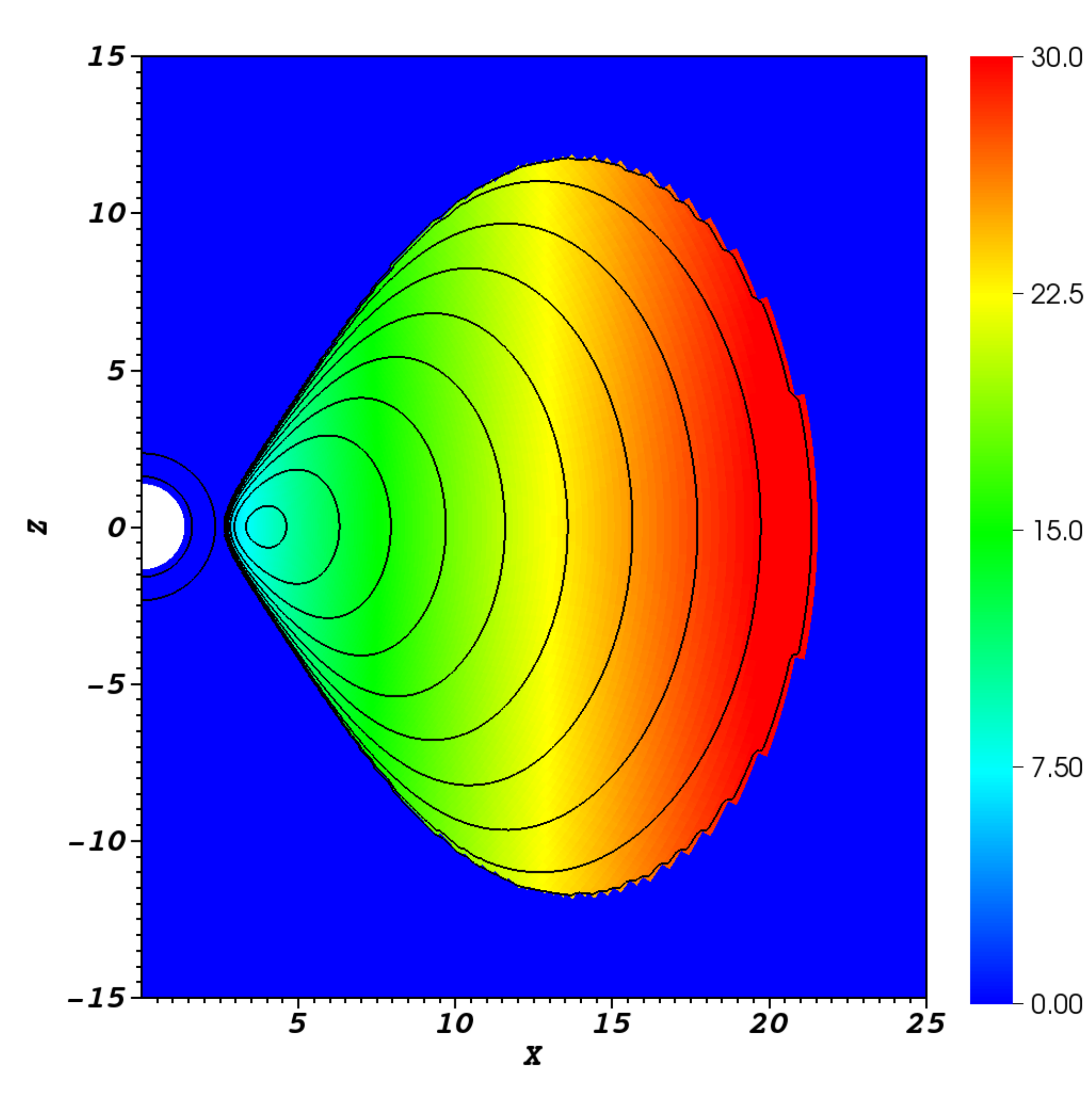}
\caption{Pseudocolor plot of the magnetization, $P_\mathrm{mag}/P_\mathrm{gas}$, with contours of gas density, $\rho$, for a poloidal slice of the strongly magnetized case (A).  The contours are spaced logarithmically over 4 orders of magnitude.}
\label{fig:initial}
\end{figure}

We consider two cases of magnetized tori, both initialized with purely toroidal magnetic fields following the procedure described in \citet{Komissarov06}.  Briefly, the Komissarov solution is uniquely specified by the following parameters:
\begin{enumerate}
\item black hole mass, $M$, and spin, $a$,
\item specific angular momentum of the torus gas, $\ell$,
\item surface potential, $W_\mathrm{in}$,
\item magnetization at the torus center, $(\pmpg)_\mathrm{c}$.
\end{enumerate}
For both our cases, the dimensionless black hole spin is $a_* = a/M = 0.9$.  The mass of the black hole is left unspecified, as it simply sets a scale for the problem.  The solutions presented in \citet{Komissarov06} all assumed a constant specific angular momentum; more general solutions with power-law angular velocities were demonstrated in \citet{Wielgus15}. In this work, we only consider constant specific angular momentum cases with $\ell = 2.8$.  While it is conceivable that one might be able to construct an initial angular momentum distribution that would alter our conclusions, we suspect that this would only be true for very unusual distributions. The remaining parameters are $W_\mathrm{in} = -0.030$ and $(\pmpg)_\mathrm{c} = 10$ (case A) or 0.1 (case Aw).  Case A comes directly from \citet{Komissarov06} and was also studied in \citet{Wielgus15}; its initial configuration is shown in Fig. \ref{fig:initial}.  Case Aw is a new case, which has parameters identical to case A, except that it has a weak initial magnetic field, $\pmpg \approx 0.1$. 

Both cases are purportedly stable according to the criteria of \citet{Pessah05}.  However, those authors only considered axisymmetric perturbations.  While the Komissarov strongly magnetized torus solution {\em is} stable to axisymmetric perturbations, it is unstable to non-axisymmetric ones, specifically the non-axisymmetric MRI \citep{Wielgus15}.

We use the {\em Cosmos++} computational astrophysics code \citep{Anninos05,Fragile12} to numerically evolve three-dimensional simulations of both tori.  This is done on a $252 \times 256 \times 64$ grid, with cells spaced evenly in coordinates $x_1$, $x_2$, and $\phi$.  $x_1$ is related to the normal radial coordinate through the logarithmic transformation, $x_1 \equiv 1 + \ln(r/r_\mathrm{BH})$, where $r_\mathrm{BH} = (1 + \sqrt{1-a_*^2}) M = 1.44 M$ is the black hole radius.  $x_2$ is related to $\theta$ as $\theta(x_2)=\pi/2 [1+(1-\varepsilon)(2x_2-1)+\varepsilon(2x_2-1)^n]$, where we take $\varepsilon = 0.7$ and $n = 29$ \citep{Noble10}.  Outflow boundaries are used at $r_\mathrm{min} = 0.95 r_\mathrm{BH}$ and $r_\mathrm{max} = 50 M$, while reflecting boundaries are employed at $\theta = 0$ and $\pi$.  We evolve only a quarter of the azimuthal domain, $0 \le \phi \le \pi/2$, with periodic boundaries at $\phi = 0$ and $\phi = \pi/2$.  We use the orbital period at $r_{c} = 4.62 M$, i.e., $t_\mathrm{orb} = 2\pi/\Omega(r_{c}) = 68 M$, as a convenient unit of time.  Both simulations are run for $\ge 10 t_\mathrm{orb}$. Note that we also performed a version of the case A simulation at half the standard resolution.  Comparison between the different resolutions suggests that our conclusions are not sensitive to the resolution used.

\section{Results}
\label{s.results}

\subsection{Weakly magnetized torus -- Case Aw}

\begin{figure}
\centering
\includegraphics[width=1\columnwidth]{\figpath/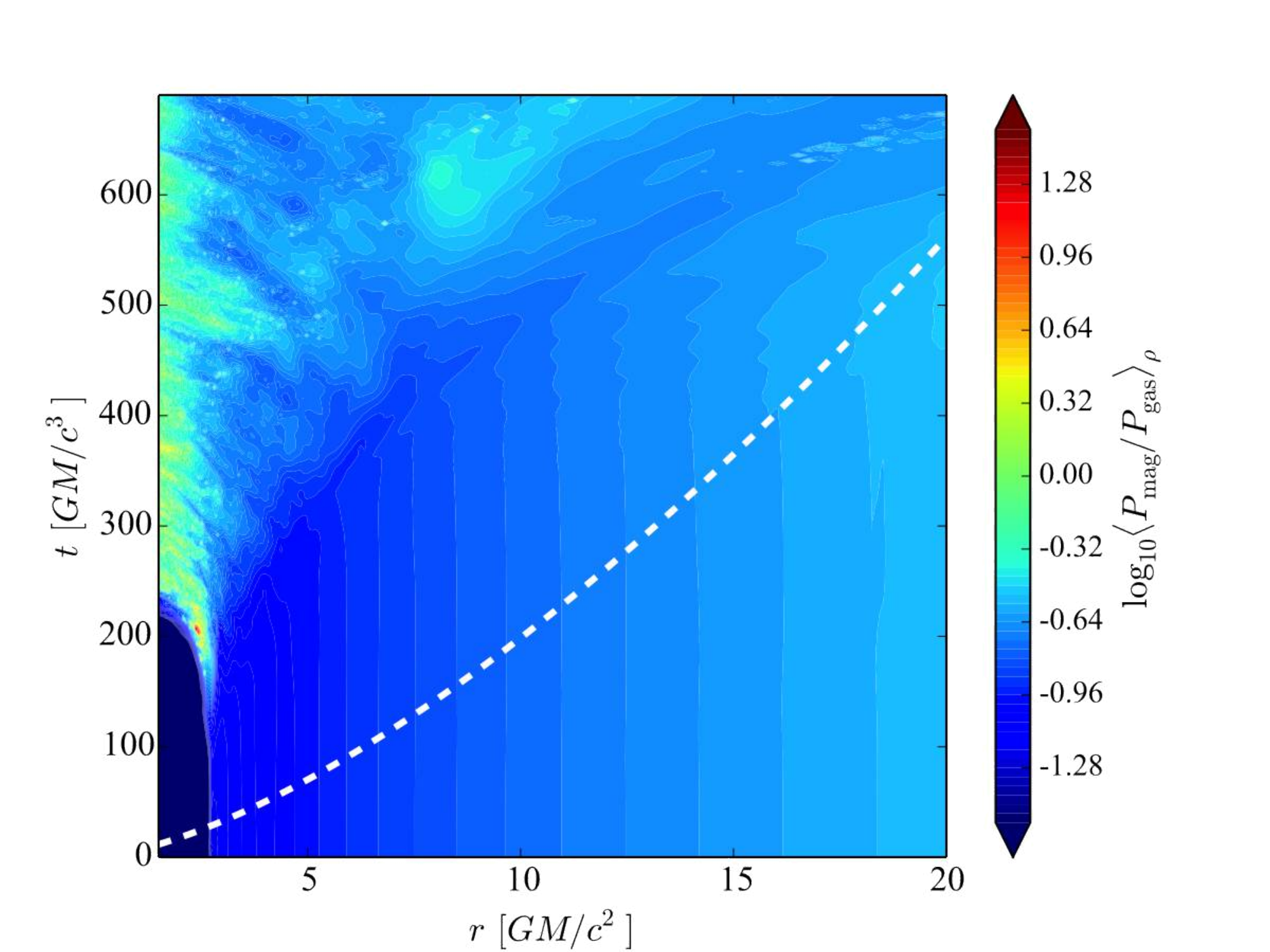}
\includegraphics[width=1\columnwidth]{\figpath/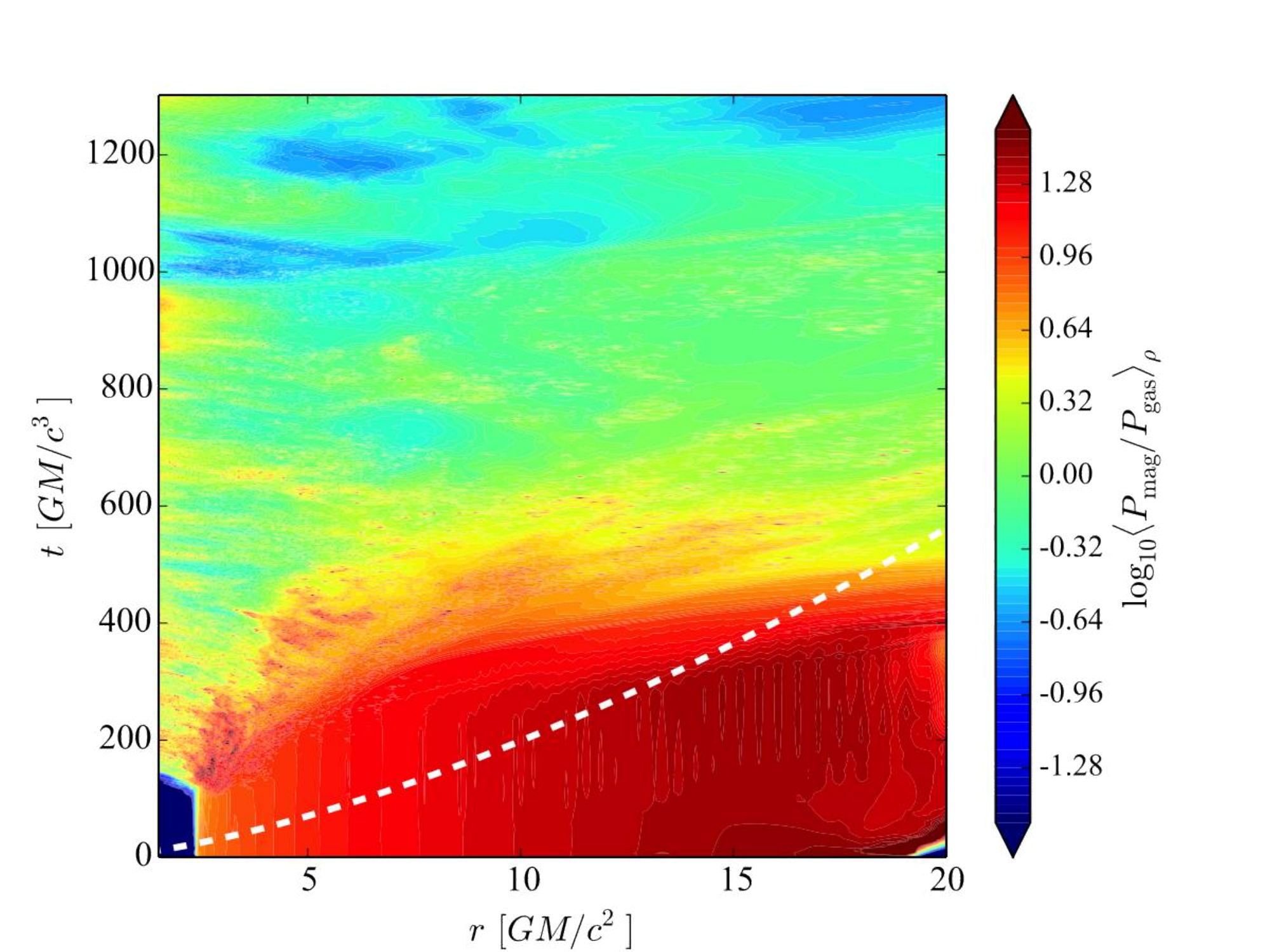}
\caption{Space-time plot of the density-weighted, shell-averaged ratio of magnetic pressure to gas pressure, $\langle P_\mathrm{mag}/P_\mathrm{gas} \rangle_\rho$, for case Aw (top panel) and case A (bottom panel). The white, dashed line shows the local orbital period for reference.}
\label{fig:pmagpgas}
\end{figure}

We start, first, by describing the results of the weakly magnetized simulation, Aw.  Consistent with numerous other simulations of weakly magnetized tori in the literature \citep[e.g.][]{Hawley00,DeVilliers03,Fragile07}, including those with weak toroidal fields \citep[e.g.][]{Beckwith08}, the magnetic field remains weak, even after the onset of the MRI.  In the top panel of Fig. \ref{fig:pmagpgas}, we plot the density-weighted, shell average of $P_\mathrm{mag}/P_\mathrm{gas}$, i.e.
\begin{equation}
\langle P_\mathrm{mag}/P_\mathrm{gas} \rangle_\rho = \frac{\int (P_\mathrm{mag}/P_\mathrm{gas}) \rho \sqrt{-g} dA_R}{\int \rho \sqrt{-g} dA_R} ~,
\label{eq:magnetization}
\end{equation}
where $\rho$ is the gas rest-mass density and $dA_R$ is the area element normal to the radial direction.  Clearly, $\langle P_\mathrm{mag}/P_\mathrm{gas} \rangle_\rho$ remains less than one and averages $\sim 0.1$ at nearly all radii and times.  

\begin{figure}
\centering
\includegraphics[width=1\columnwidth]{\figpath/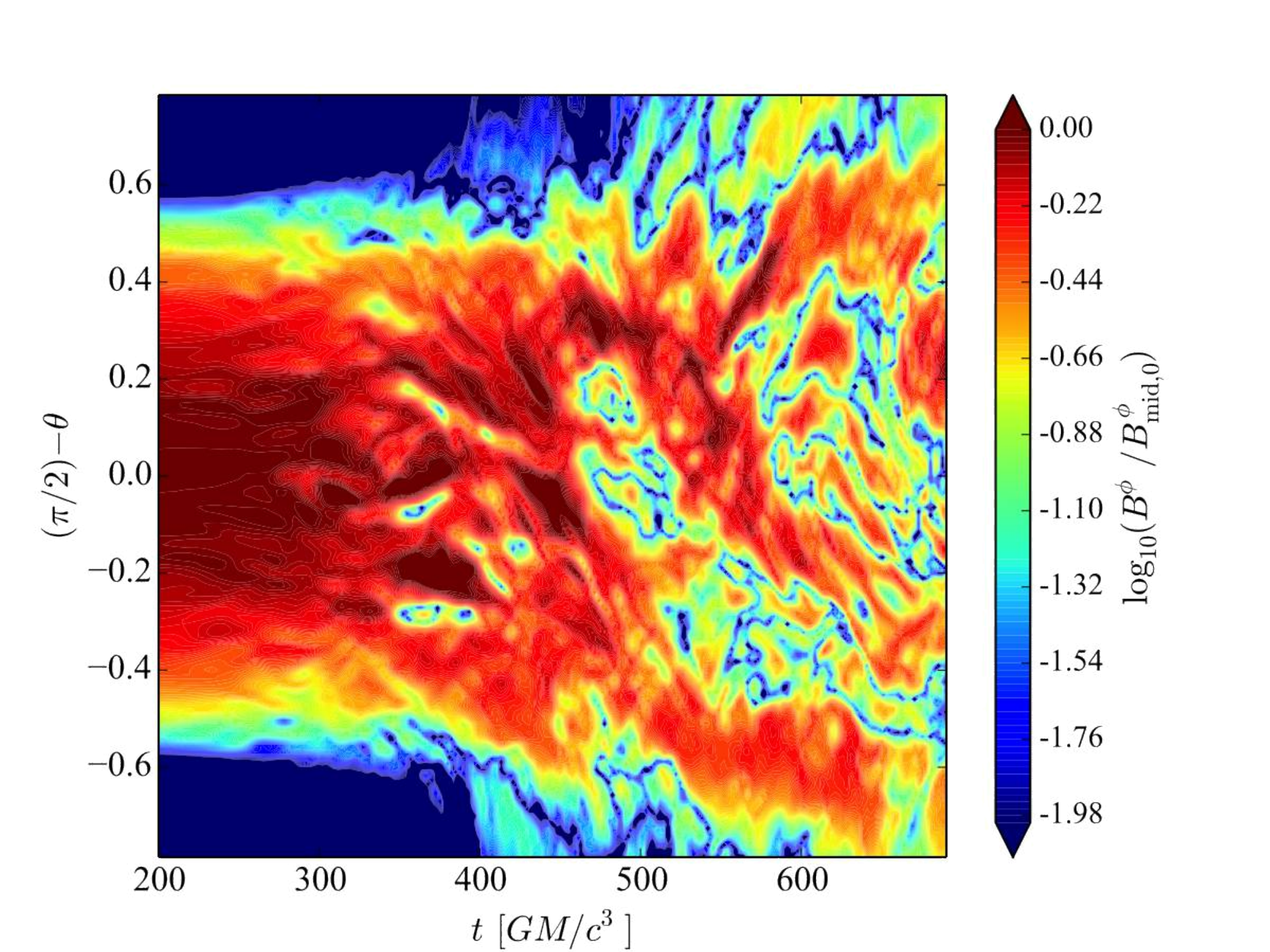}
\includegraphics[width=1\columnwidth]{\figpath/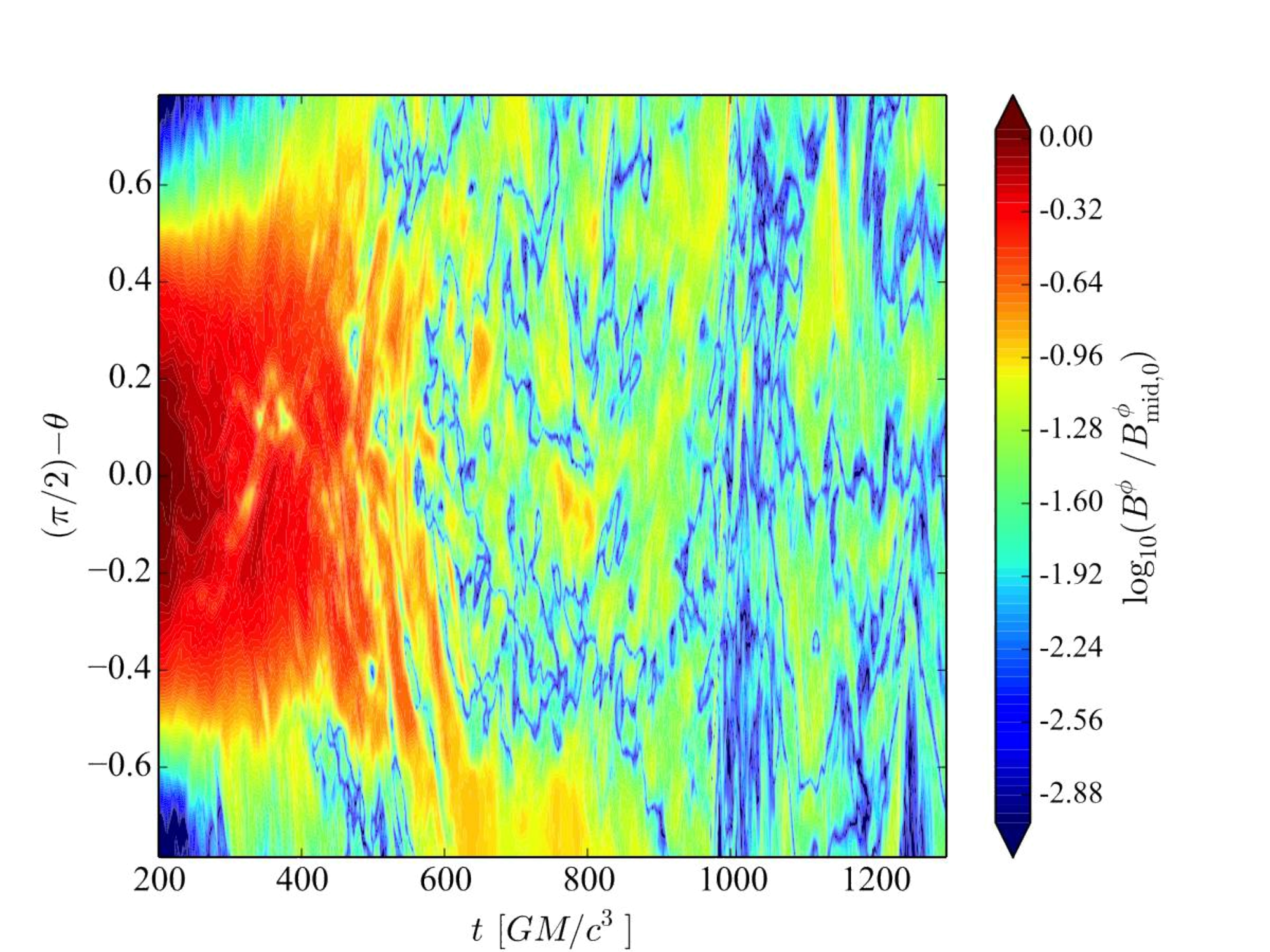}
\caption{Space-time diagram of the azimuthally averaged, azimuthal magnetic field, $B^\phi$, at $r = r_\mathrm{c}$, normalized by its initial mid-plane value for case Aw (top panel) and case A (bottom panel).  Note the different scales on the two plots.  In case A, the magnetic field becomes significantly weaker with time.}
\label{fig:butterfly}
\end{figure}

Fig. \ref{fig:butterfly} presents a so-called ``butterfly'' diagram, which is intended to show the cyclic decay and replenishment of the azimuthal magnetic field owing to the action of the MRI dynamo.  In shearing box simulations with weak or no net vertical flux, such diagrams show very regular field reversals about every 10 orbits \citep{Brandenburg95,Davis10,Simon12}, producing the colorful pattern that gives these diagrams their name.  However, the pattern is not always so orderly.  In shearing box simulations with significant net vertical flux, the behavior becomes more chaotic, with the dynamo behavior ceasing altogether above a certain field strength \citep{Fromang12,Bai13,Salvesen16a}.  For our weakly magnetized case Aw (top panel), the behavior of $B^\phi$ is quite sporadic.  The field initially decays, but is replenished to roughly its initial strength repeatedly and in multiple locations, though no clear periodic behavior is seen.  Importantly, $B^\phi$ does not grow significantly stronger, consistent with the magnetization remaining weak.

\begin{figure}
\centering
\includegraphics[width=1\columnwidth]{\figpath/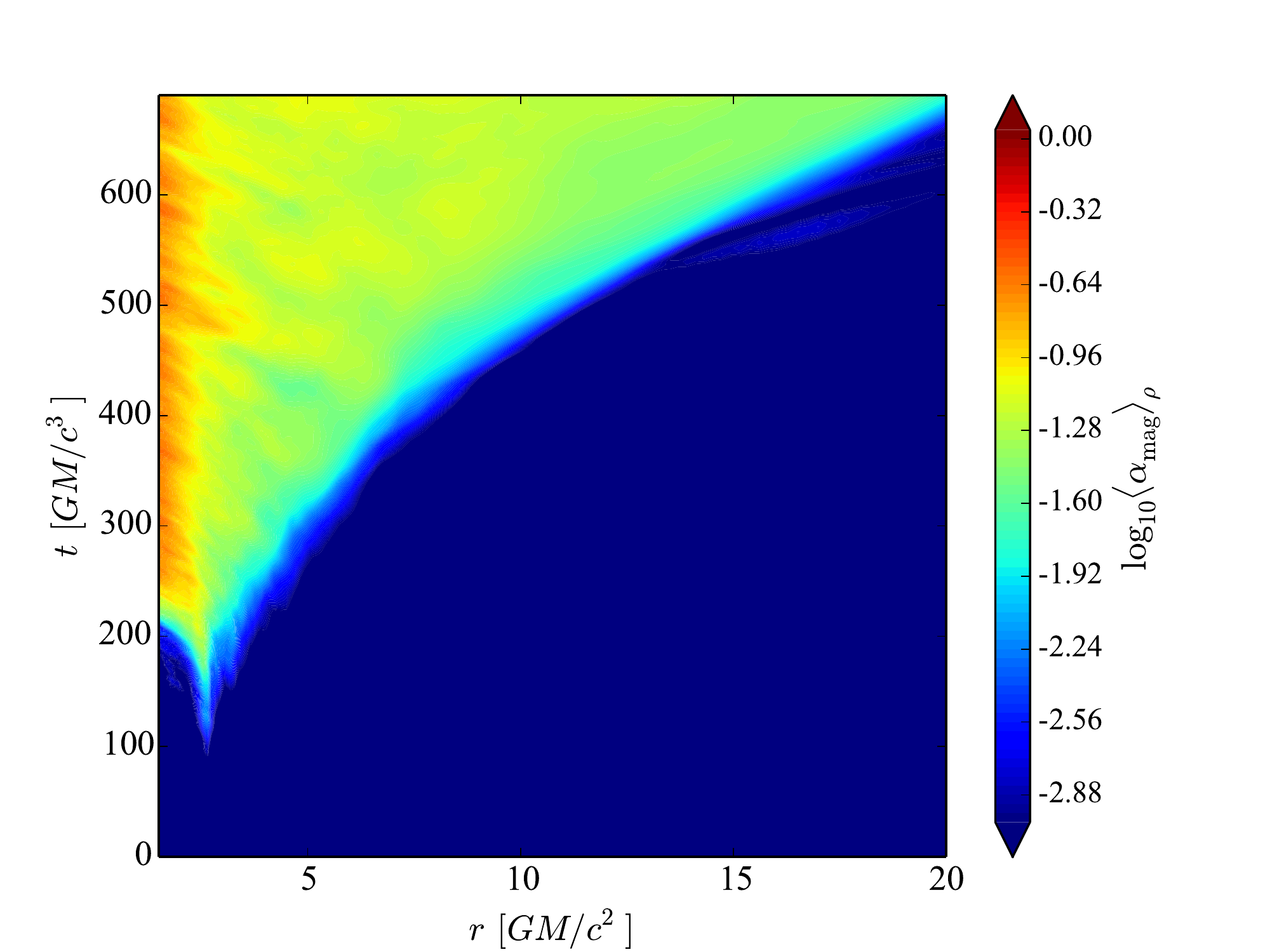}
\includegraphics[width=1\columnwidth]{\figpath/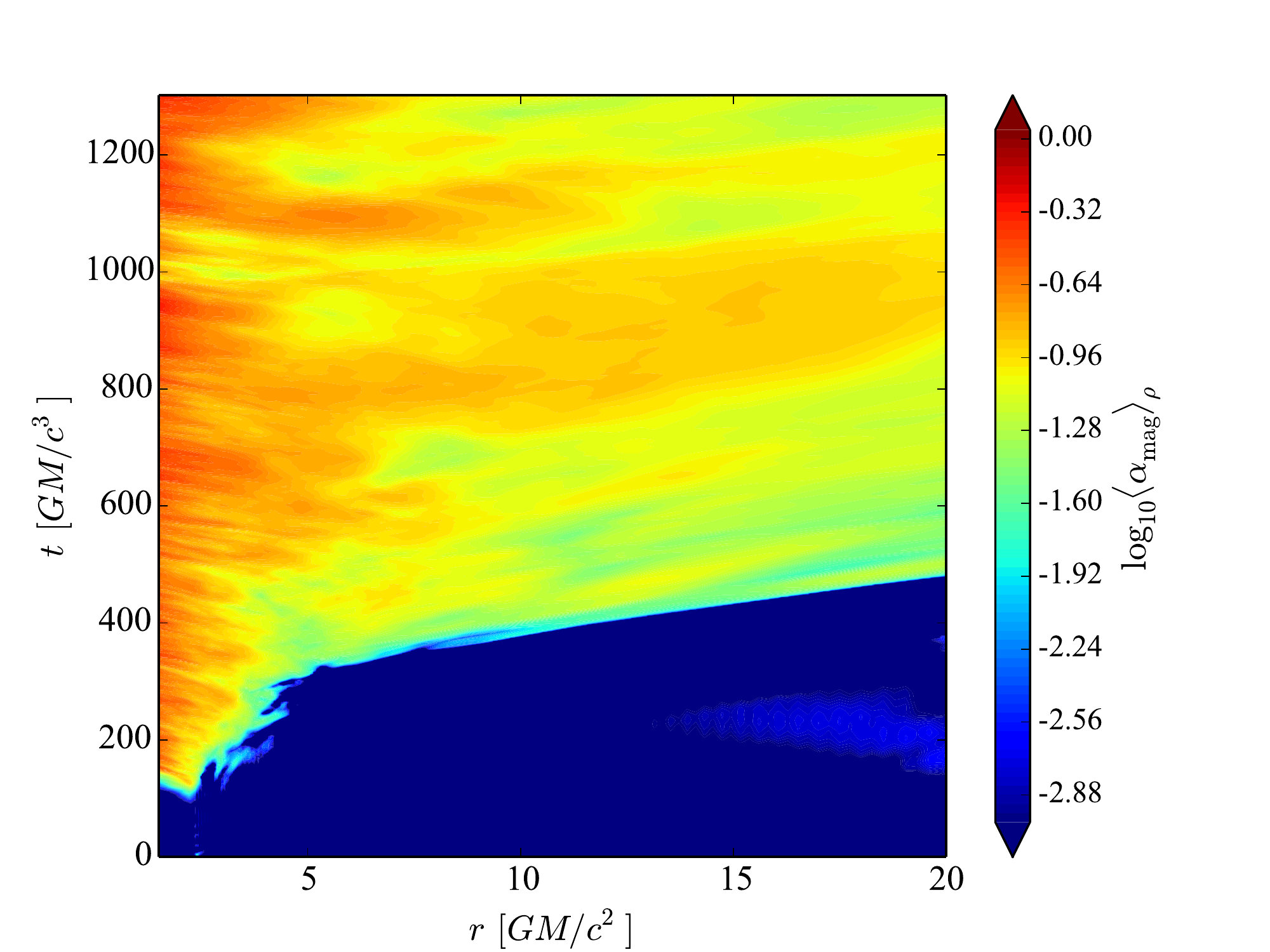}
\caption{Space-time plot of the density-weighted, shell-averaged $\alpha_\mathrm{mag}$ for case Aw (top panel) and case A (bottom panel).}
\label{fig:alpha}
\end{figure}

Finally, in Fig. \ref{fig:alpha}, we consider the dimensionless stress parameter
\begin{equation}
\alpha_\mathrm{mag} = \frac{(W_{\hat{r}\hat{\phi}})_\mathrm{mag}}{P_\mathrm{Tot}} ~,
\end{equation}
averaged and weighted in the same manner as the magnetization in equation (\ref{eq:magnetization}), where $W_{\hat{r}\hat{\phi}}$ is the covariant $\hat{r}$-$\hat{\phi}$ component of the magnetic stress tensor and $P_\mathrm{Tot} = P_\mathrm{mag} + P_\mathrm{gas}$ represents the total pressure.  The top panel (representing case Aw) shows a fairly typical value for a weakly magnetized disc simulation ($\alpha_\mathrm{mag} \approx 0.02$).  

\subsection{Strongly magnetized torus -- Case A}

\begin{figure}
\centering
\includegraphics[width=1\columnwidth]{\figpath/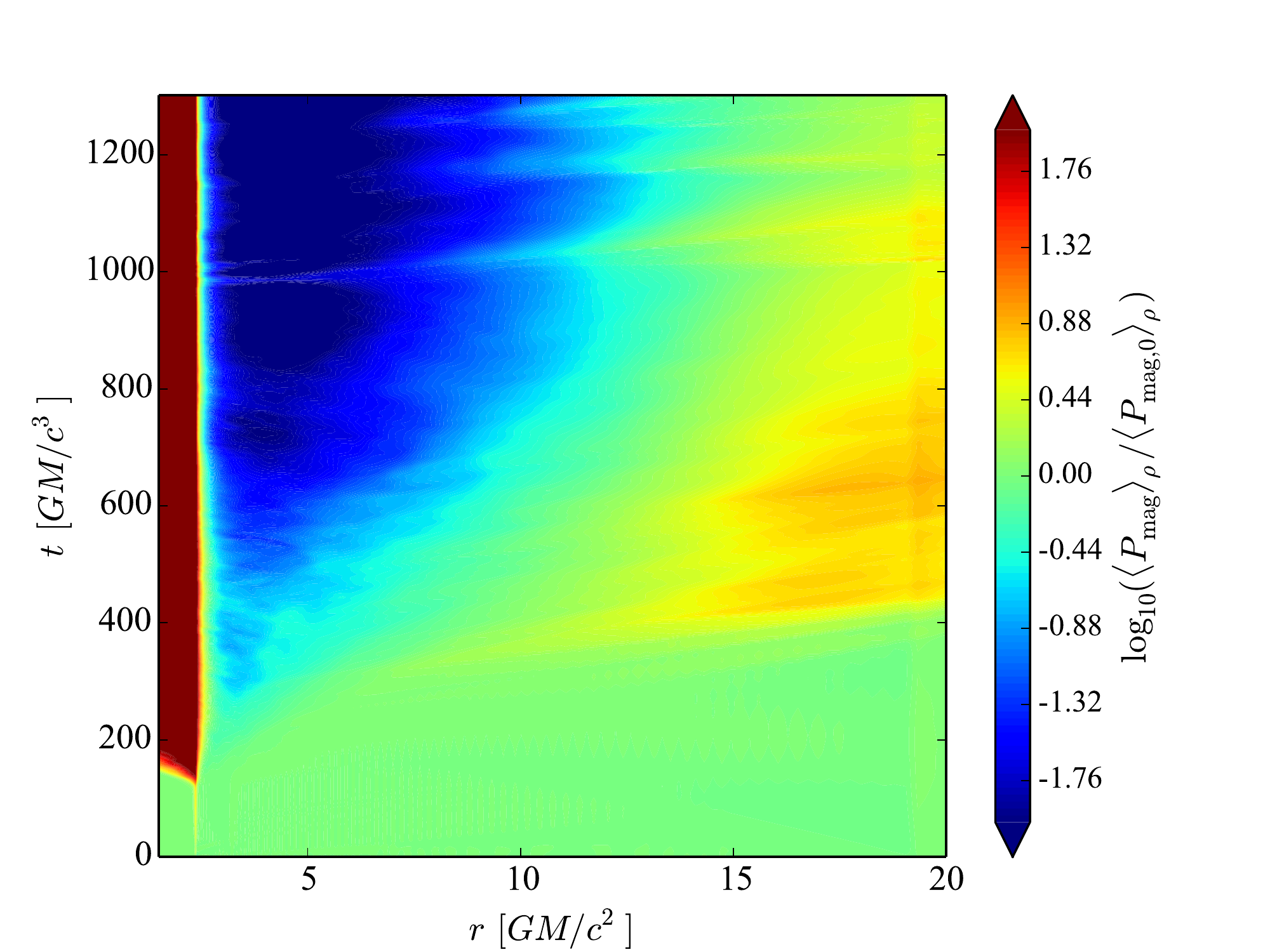}
\includegraphics[width=1\columnwidth]{\figpath/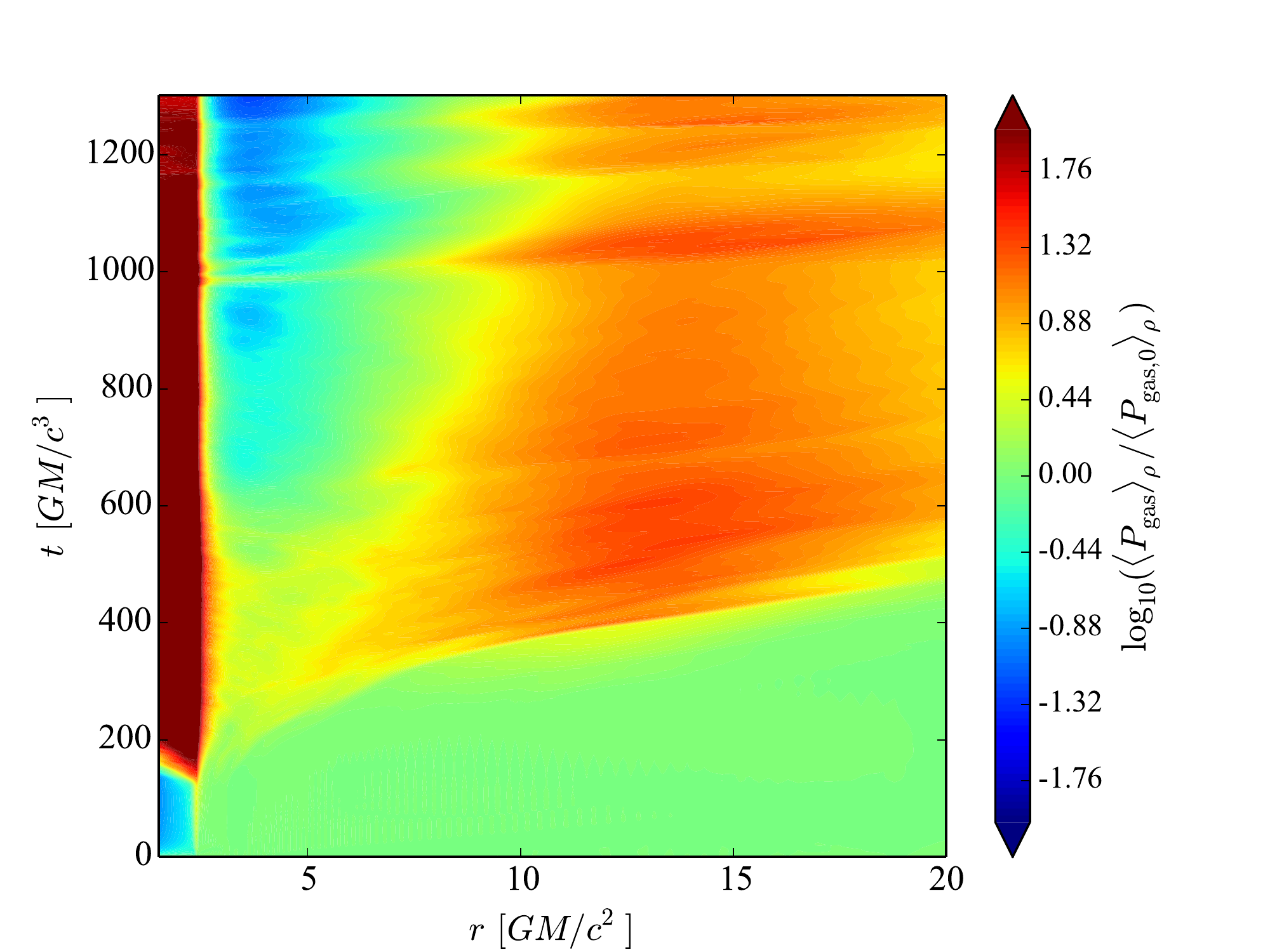}
\caption{Top panel: Space-time plot of the density-weighted, shell-averaged value of magnetic pressure, $\langle P_\mathrm{mag} \rangle_\rho$, normalized by its initial value, for case A.  Bottom panel: Similar space-time plot, but for the density-weighted, shell-averaged value of gas pressure, $\langle P_\mathrm{gas} \rangle_\rho$, normalized by its initial value.}
\label{fig:pmag}
\end{figure}

The results in the previous section were all consistent with previous global simulations of weakly magnetized tori.  The results of this section will be more novel.  First off, we claim that strong toroidal magnetization can not be maintained within a disc for more than $\sim 10$ orbital periods.  The strongest evidence for this is presented in Figs. \ref{fig:pmagpgas} (bottom panel) and \ref{fig:pmag}.  Fig. \ref{fig:pmagpgas} clearly shows that the magnetization within the disc drops steadily from $\langle P_\mathrm{mag}/P_\mathrm{gas} \rangle_\rho \ge 10$ to $\lesssim 0.1$, two orders of magnitude in roughly 20 orbital periods.  Fig. \ref{fig:pmag} shows that, at small radii ($r \lesssim 10 M$), the drop in $\langle \pmpg \rangle_\rho$ is owing almost entirely to a very significant drop in $\langle \pm \rangle_\rho$.  At larger radii ($r \gtrsim 10 M$), the drop in $\langle \pmpg \rangle_\rho$ is more attributable to a rise in $\langle \pg \rangle_\rho$.

\begin{figure}
\centering
\includegraphics[width=0.9\columnwidth]{\figpath/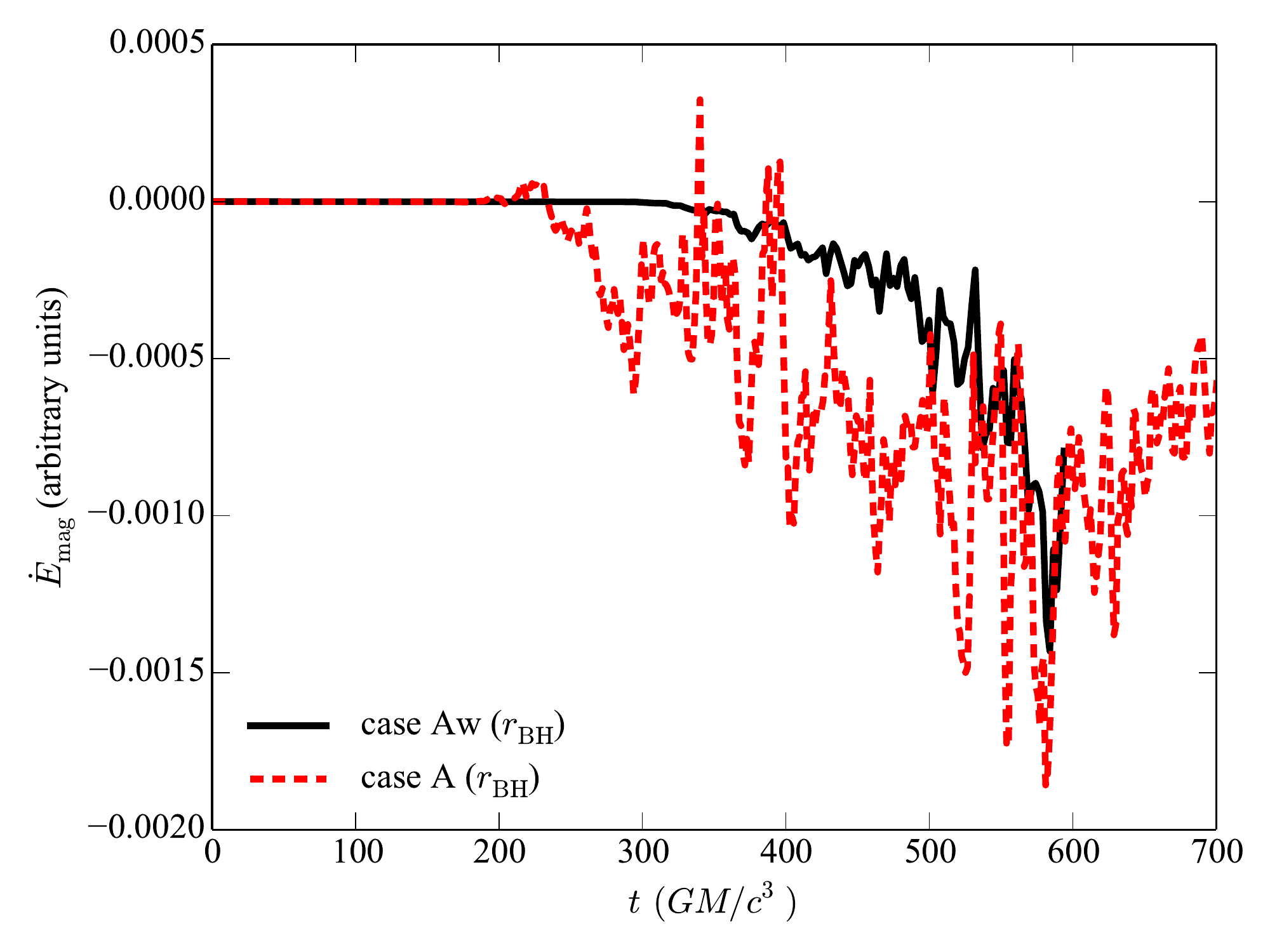}
\includegraphics[width=0.9\columnwidth]{\figpath/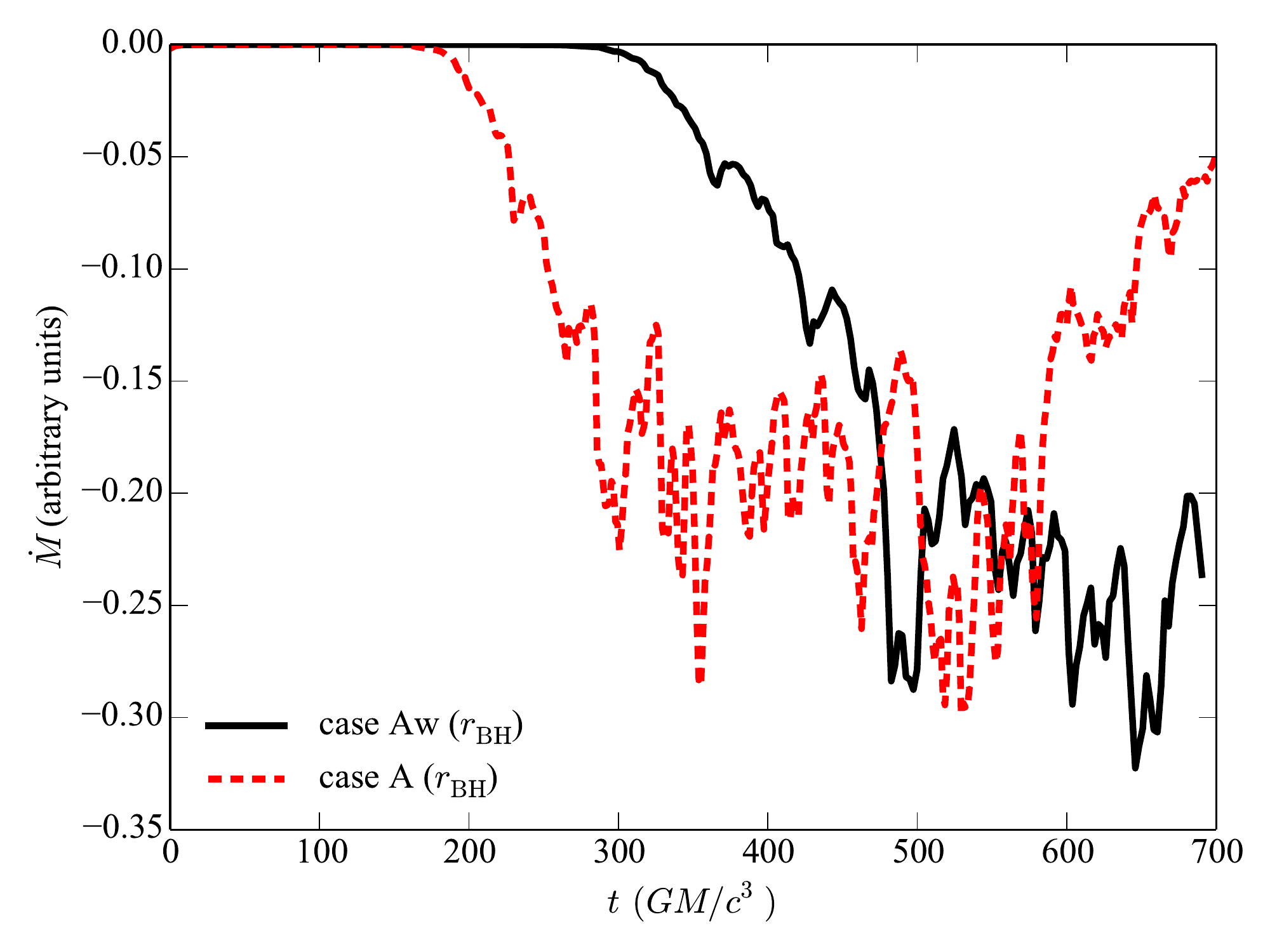} 
\includegraphics[width=0.9\columnwidth]{\figpath/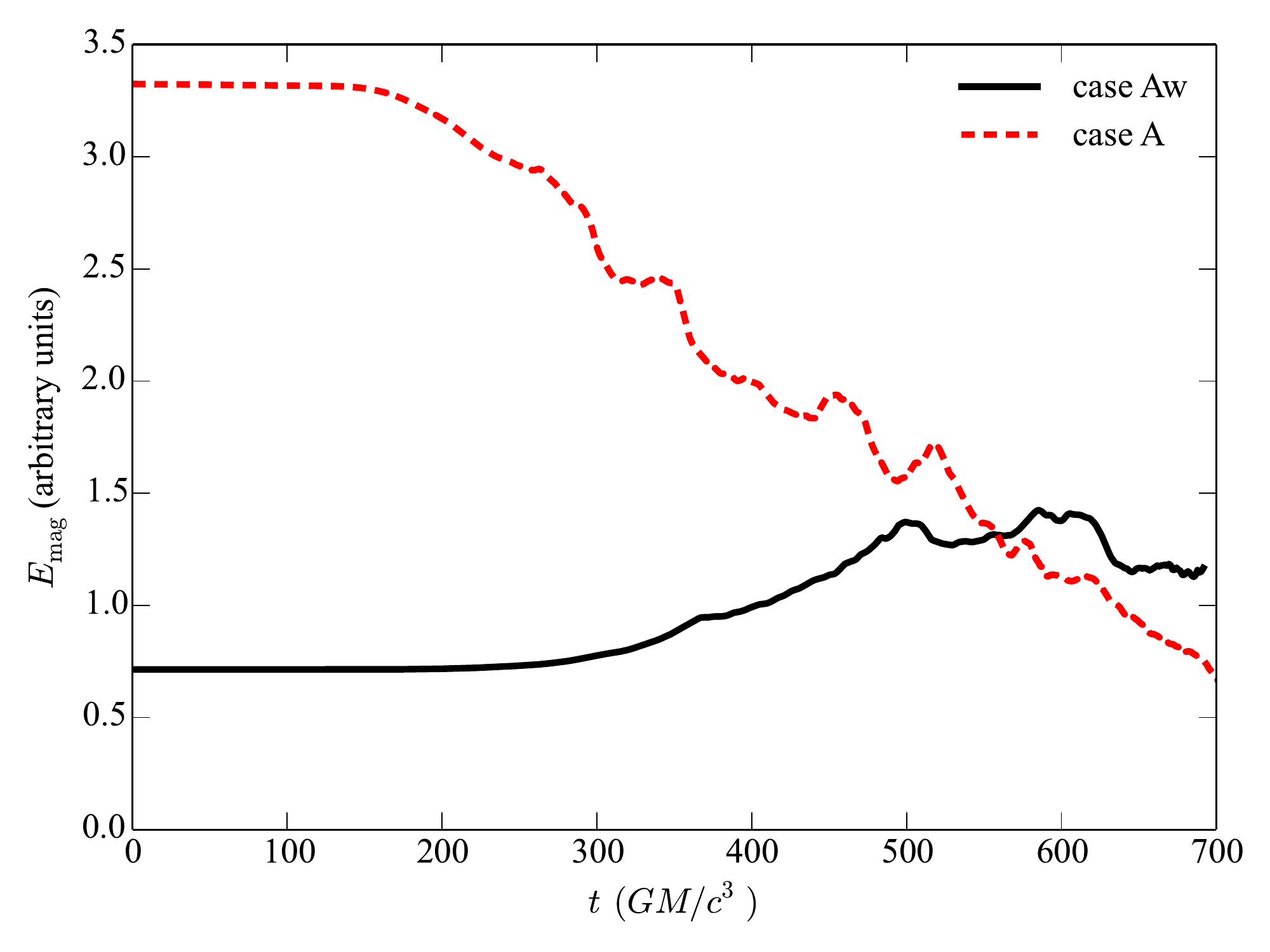}
\includegraphics[width=0.9\columnwidth]{\figpath/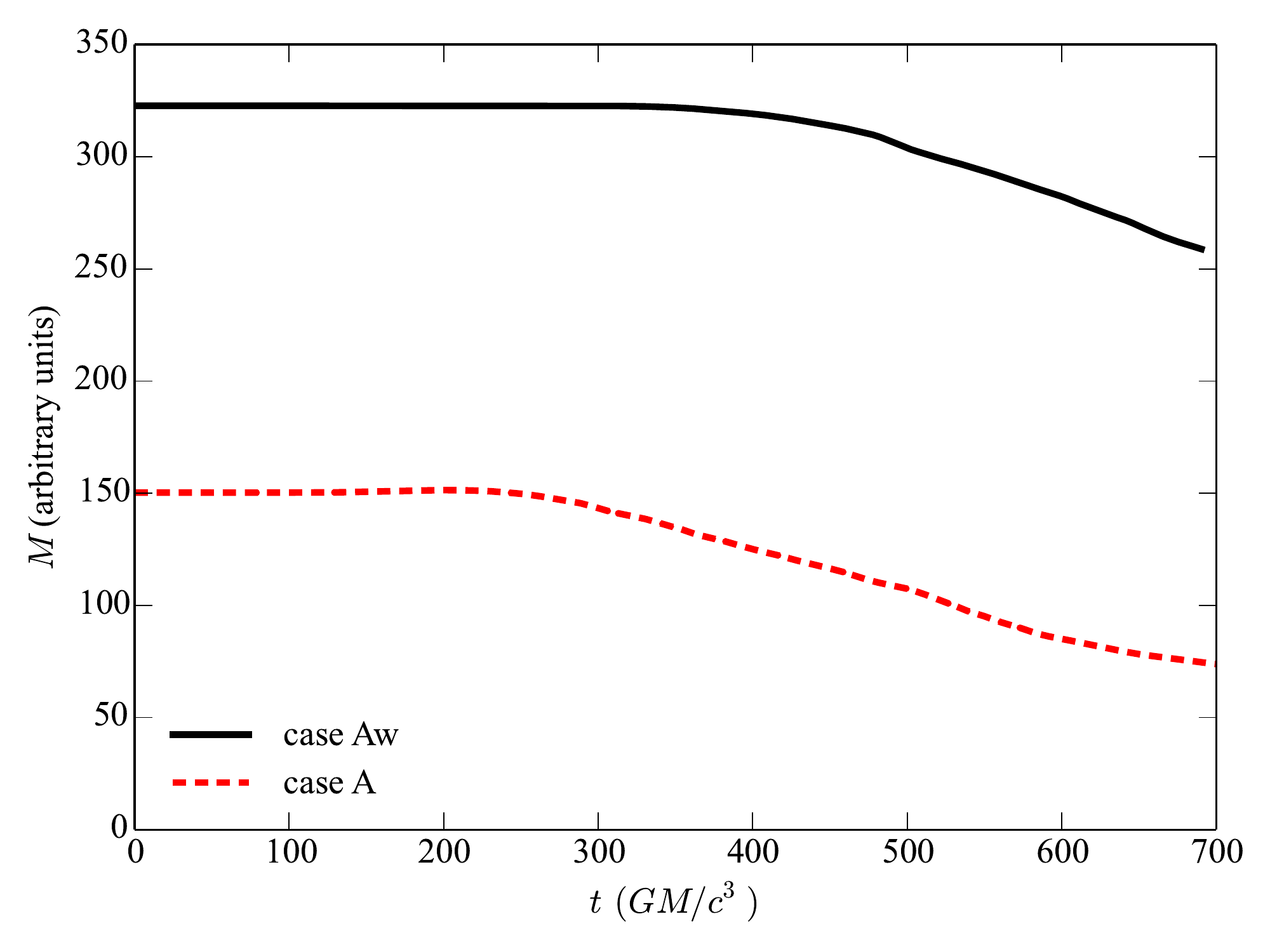}
\caption{Magnetic (top) and mass (second) fluxes at the event horizon and total magnetic energy (third) and mass (bottom), all in arbitrary units, as a function of time for cases Aw and A.}
\label{fig:fluxes}
\end{figure}

Our first consideration is that the drop in  $\langle \pm \rangle_\rho$ at small radii may be due to advection of the magnetic field into the black hole.  In Figure \ref{fig:fluxes}, we compare the magnetic flux at the horizon, i.e.
\begin{equation}
\dot{E}_\mathrm{mag}(t) = \int \sqrt{-g} (T^r_t)_\mathrm{mag} dA_R ~,
\label{eq:mag_flux}
\end{equation}
with the mass flux, $\dot{M}(t)$ (calculated by replacing $(T^r_t)_\mathrm{mag}$ in equation (\ref{eq:mag_flux}) with $\rho u^r$). We also show the total integrated magnetic energy 
\begin{equation}
E_\mathrm{mag}(t) = -\int \sqrt{-g} (T^t_t)_\mathrm{mag} dV
\label{eq:total_mag}
\end{equation}
and mass, $M$ (calculated by replacing $(T^t_t)_\mathrm{mag}$ in equation (\ref{eq:total_mag}) with $-\rho u^t$), on the grid as a function of time.  We find that the flux of magnetic field into the black hole is insignificant when compared to the total drop in magnetic energy on the grid.  This is unlike the mass, where the flux at the event horizon nicely matches the drop in total mass, as it should. (Magnetic energy losses through the outer radial boundary are negligible in both cases.)  Therefore, the significant drop in magnetic energy must have another cause.  

Reconnection is another possibility to consider.  In many MRI turbulent disc simulations, magnetic energy is converted to gas internal energy (heat) through reconnection at the grid scale.  However, given that our global field topology is purely azimuthal (and in a uniform direction) and reconnection within the computational domain can not change this topology, it is not possible to lower the overall magnetic energy in this way.  Reconnection can, and certainly does, take place during the MRI-driven turbulence phase, but it can not, by itself, be responsible for the reduction of total magnetic energy that we see in case A. 

\begin{figure}
\centering
\includegraphics[width=1\columnwidth]{\figpath/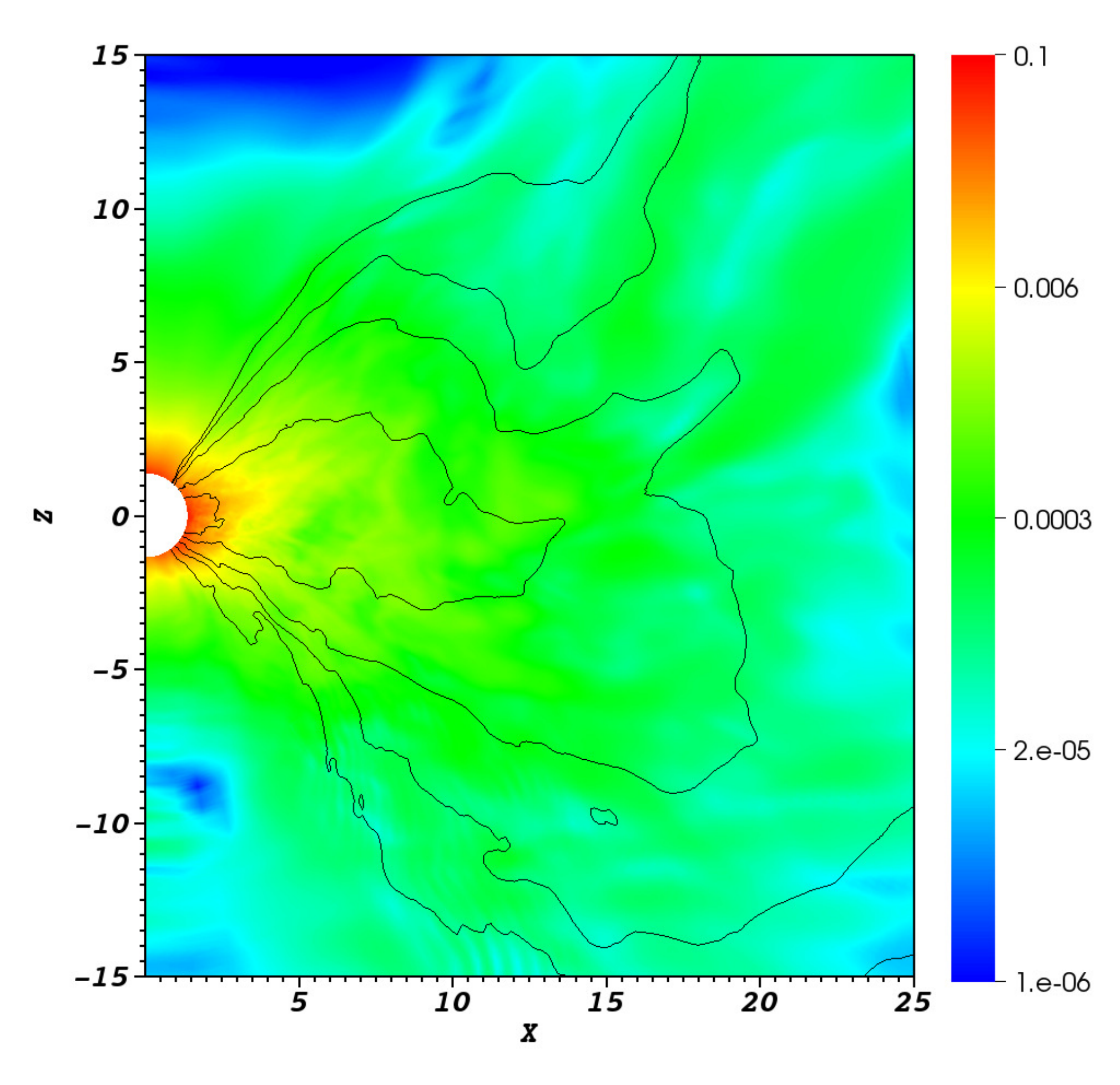}
\caption{Pseudocolor plot of the magnetic pressure normalized to its initial maximum value, $P_\mathrm{mag}/(P_\mathrm{mag})_{0,\mathrm{max}}$, with contours of gas density, $\rho$, for the strongly magnetized case (A).  The contours are spaced logarithmically over 4 orders of magnitude.  Data are time averaged over the final orbital period, $t_\mathrm{orb} = 68 M$, of the simulation.}
\label{fig:pmag_rho}
\end{figure}

It seems, in the end, that most of the drop in magnetic energy in Figure \ref{fig:fluxes} is owing to work being done by the magnetic field on the gas.  Whereas in weakly magnetized discs the MRI dynamo often increases the total magnetic energy (see case Aw in the lower left panel of Figure \ref{fig:fluxes}), in the strongly magnetized case, it seems the MRI facilitates a redistribution of the gas and magnetic field, such that the magnetization drops and the disc becomes weakly magnetized in its interior. As a result of this redistribution, the magnetic field that initially occupied the inner disc region is dispersed over a much larger volume by the end of the simulation.  This can be seen in Fig. \ref{fig:pmag_rho}, where much of the field is found outside the disc, especially in the funnel region.  This spreading of the field over a larger volume necessarily results in a drop in magnetic pressure within the disc.

Additional insight can be gained by looking again at Fig. \ref{fig:butterfly}.  As before, we see considerable variability in the strength of $B^\phi$ for case A (bottom panel), but no well-organized cycle.  Furthermore, in this case, the field fails to return to anything close to its initial value.  Instead, the mean field strength is reduced by more than an order of magnitude.  This suggests that, similar to net poloidal flux in shearing box simulations, strong toroidal fields in global simulations can interfere with the MRI dynamo.  There is also no evidence for a Parker-driven dynamo.

One final note of interest regarding the strongly magnetized torus case is that, for the brief period that the magnetization, $\pmpg$, is high, the dimensionless stress, $\alpha_\mathrm{mag}$, is also high (compare Figs. \ref{fig:pmagpgas} and \ref{fig:alpha}).  \citet{Bai13,Salvesen16a} showed that discs threaded with a net vertical flux exhibit systematically higher values of $\alpha$ as the flux increases, but this is the first evidence that we are aware of a similar dependence of $\alpha_\mathrm{mag}$ on the strength of the toroidal field.  For our finite size torus, where the disc never reaches a true steady state, this makes a noticeable difference in how quickly material accretes onto the black hole (Figure \ref{fig:fluxes} right panels).  However, this appears to be a transient effect in our case.  If we had run a longer simulation by either starting with a larger torus or one located further from the black hole, then we anticipate that the value for $\alpha_\mathrm{mag}$ would have steadily declined, as did the magnetization, to a value comparable to our weak magnetization case.

\section{Discussion and Conclusions}
\label{s.discuss}

In this work, we studied the evolution of magnetic fields in initially strongly magnetized (toroidal field) global simulations, similar in nature to shearing box simulations of \citet{Johansen08} and \citet{Salvesen16}.  We started with magnetic fields of differing strengths [$(\pmpg)_\mathrm{c} = 0.1$ or 10] and let the simulations run for many orbital periods.  In both cases, we found that the disc ultimately reached weakly magnetized configurations ($\langle \pmpg \rangle_\rho \sim 0.1$).  This same limit has been reached in all weak initial field configurations of which we are aware \citep[see][and references therein]{Hawley11,Hawley13} and also for the strong, toroidal fields considered in \citet{Salvesen16}.  Only when large scale (i.e. non-local) fields are present, as in the case of a net poloidal flux \citep{Bai13,Salvesen16a} or strong radial component \citep{Sadowski16}, are strong magnetizations sustained within the disc.  

In our initially high (toroidal) magnetization case, we find that the magnetization drops mostly because of a redistribution of the gas and magnetic fields.  The strong magnetic fields do work against the gas, pushing their way into the funnel region and out to large radii. This happens over a few tens of orbital periods, and the azimuthal field in the disc is never fully replenished.  Thus, this simulation shows no evidence of the Parker-driven dynamo predicted by \citet{Johansen08}.

Our work confirms the shearing box results of \citet{Salvesen16} that discs with predominantly toroidal magnetic fields can not remain strongly magnetized ($\pmpg > 1$), at least not within 1-2 scale heights of the disc midplane.  In other words, strong toroidal magnetic fields are not locally self-sustaining within the body of the disc.  If strongly magnetized discs are then required to explain the apparent stability of many accreting black hole X-ray binaries at high luminosities, then the discs must be threaded by some other dominant field component, either radial or vertical.  Additionally, the thermal collapse scenario for creating strongly magnetized discs, even if it is possible, must lead to only a temporary increase in magnetization, lasting only a few tens of orbits.

\section{Acknowledgements}

This research was supported by the National Science Foundation under grant NSF AST-1211230 and AST-1616185.  This work used the Extreme Science and Engineering Discovery Environment (XSEDE), which is supported by National Science Foundation grant number ACI-1053575.  We also thank the International Space Science Institute, where part of this work was carried out, for their hospitality.

\bibliographystyle{mn2e}

\end{document}